\documentclass[%
aps,
superscriptaddress,
nofootinbib,
amsmath,amssymb,
aps,
prb,
twocolumn,
]{revtex4-2}

\usepackage{dcolumn}
\usepackage{bm}
\usepackage{graphicx,color}
\usepackage[english]{babel}
\usepackage[colorlinks]{hyperref}
\usepackage{comment}
\usepackage{tabularx}
\usepackage{array}
\usepackage[svgnames,table]{xcolor}
\usepackage{ragged2e}

\newcolumntype{A}{>{\centering\arraybackslash \columncolor{white!50!white}}m{2.1cm}}
\newcolumntype{B}{>{\centering\arraybackslash \columncolor{white}}m{7.9cm}}
\newcolumntype{C}{>{\centering\arraybackslash \columncolor{white!50}}m{7.9cm}}
\newcolumntype{D}{>{\centering\arraybackslash \columncolor{white!42}}m}
\newcolumntype{P}[1]{>{\centering\arraybackslash}p{#1}}
\def\beq{\begin{equation}}
	\def\eeq{\end{equation}}
\def\bea{\begin{eqnarray}}
	\def\eea{\end{eqnarray}}
\def\barr{\begin{array}}
	\def\earr{\end{array}}

\usepackage{caption}
\usepackage{color}
\usepackage[font=Large]{subfig}
\usepackage{tabularx,ragged2e,booktabs,caption}
\usepackage{grffile}

\def\ba{\begin{eqnarray}}
	\def\ea{\end{eqnarray}}

\begin{document}
	
	\preprint{APS/123-QED}
	
	\title{The effect of screening on the relaxation dynamics in the Coulomb glass}
	
	\author{Preeti Bhandari}
	\affiliation{Department of Physics, Ben-Gurion University of the Negev, Beer Sheva 84105, Israel.}
	\author{Vikas Malik}
	\email{vikasm76@gmail.com}
	\affiliation{Department of Physics and Material Science, Jaypee Institute of Information Technology, Uttar Pradesh 201309, India.}
	\author{Moshe Schechter}%
	\affiliation{
		Department of Physics, Ben-Gurion University of the Negev, Beer Sheva 84105, Israel
	}%
	
	\date{\today}
	
	\begin{abstract}
		This paper examines the relaxation dynamics of a two-dimensional Coulomb glass lattice model with high disorders. The study aims to investigate the effects of disorder and Coulomb interactions on glassy dynamics by computing the eigenvalue distribution of the linear dynamical matrix using mean-field approximations. The findings highlight the significance of the single-particle density of states (DOS) as the main controlling parameter affecting the relaxation at intermediate and long times. For the model with unscreened Coulomb interactions, our results indicate that the depletion of the DOS near the Fermi level leads to logarithmic decay at intermediate times. As the relaxation progresses to longer times, a power-law decay emerges, with the exponent approaching zero as the disorder strength increases, suggesting the manifestation of logarithmic decay at high disorders. The effects of screening of interactions on the dynamics are also studied at various screening and disorder strengths. The findings reveal that screening leads to the filling of the gap in the density of states, causing deviation from logarithmic decay at intermediate disorders. Moreover, in the strong disorder regime, the relaxation dynamics are dominated by disorder, and even with screened Coulomb interactions, the electronic relaxation remains similar to the unscreened case. The time at which crossover to exponential decay occurs increases with increasing disorder and interaction strength.
	\end{abstract}
	
	
	\maketitle
	
	\section{Introduction}
	\label{intro} 
	
	Slow dynamics is one of the most striking features of glasses, as observed both numerically \cite{bhandari2021relaxation,kolton2005heterogeneous,kirkengen2009slow,burin2006many} and experimentally \cite{cugliandolo1993analytical,cugliandolo1994out,vaknin2000aging,vaknin2000heuristic,vaknin2002nonequilibrium,orlyanchik2004stress,kozub2008memory,ben1993nonequilibrium,martinez1997anomalous,martinez1998coulomb}. Understanding the origin of these slow dynamics is an important problem in condensed matter physics. In disordered electronic systems, it is generally believed that the interplay of disorder and unscreened Coulomb interaction results in glassy behavior. The Coulomb Glass (CG) model, which exhibits many characteristics of glass \cite{davies1982electron,grunewald1982mean,pollak1982coulomb,pollak1984non,grannan1994grannan,mueller2004glass,pastor1999melting,pankov2005nonlinear,mueller2007mean,bray1982spin,amir2009slow,amir2008mean,burin2006many}, provides an excellent framework for understanding these phenomena. The CG model describes a disordered lattice of electrons that interact via unscreened Coulomb interactions. The strength of disorder and interaction between the electrons play an important role in the formation of the soft Coulomb gap at high disorders \cite{efros1975coulomb,shklovskii2013electronic,baranovskii1979coulomb,davies1984properties,mobius1992coulomb,glatz2008coulomb,bhandari2017critical,bhandari2017effect,goethe2009phase,surer2009density,mobius2010comment}. The gap in the single-particle density of states (DOS) of the system is filled up as the temperature is increased \cite{sarvestani1995coulomb}, or if the electron-electron interaction is screened.  Since unscreened Coulomb interactions are pivotal to the formation of the soft Coulomb gap at high disorder, one concludes that the slow relaxation is due to the interplay between disorder and interactions. This has been observed experimentally \cite{ovadyahu1997disorder,vaknin1998evidence,ovadyahu2017slow,ovadyahu2018transition,orlyanchik2007electron,delahaye2020electron,pollak2006model,sagui1994spinodal,maheshwari1993morphological} for samples where both disorder and interactions are strong, but the question remains about the role of long-range Coulomb interactions played in slow relaxation.
	
	The relaxation dynamics in a CG system can be studied experimentally in a variety of procedures. For example, quenching the system from high temperatures to low temperatures \cite{ovadyahu2003history}. In this case, one observes that the excess conductance of the sample initially relaxes very fast, followed by a slow relaxation. Similarly, a non-equilibrium state can also be created using gate protocols \cite{vaknin2000aging,grenet2003symmetrical} or by absorption of light \cite{ovadyahu2015infrared,ovadyahu2009conductance}. In all cases, the slow relaxation behavior can be explained by the formation of the Coulomb gap in the density of states (DOS) \cite{clare1999time,malik2004formation,lebanon2005memory}. The gap forms slowly with time, and its width depends on the strength of the disorder and the electron-electron interactions.
	
	Experiments \cite{ovadyahu2019screening} have also been carried out on samples having screened Coulomb interactions, in which a metallic plate is employed to screen the interaction between electrons. The sluggish dynamics seen in these samples are surprisingly quite similar to those in the reference sample without the metallic plate. 
	
	In this paper, we investigate the role of screening on slow dynamics in the CG model using mean-field approximation. We compare the dynamics with unscreened Coulomb interactions to the dynamics with screened Coulomb interactions as a function of disorder strength. Our aim is to gain a better understanding of the interplay between disorder and interactions and the role of screening on slow dynamics. Other effects of the screening not considered here, such as the polaronic effect \cite{asban2021polaronic}, may affect the dynamics.
	
	The Hamiltonian of a CG lattice model has been defined in terms of occupation numbers $n_i$ and the on-site random field energy $\phi_i$. In dimensionless units, the Hamiltonian \cite{efros1976coulomb,efros2011coulomb,mobius2009coulomb,pikus1994critical} is given by
	
	\begin{equation}
		\label{Hamiltonian}
		\mathcal{H} \{n_{i}\} = \sum_{i=1}^{N} \phi_{i} n_{i} + \frac{1}{2} \sum_{i \neq j} \frac{e^{2}}{\kappa |\vec{r_{i}} - \vec{r_{j}}|} (n_{i} - 1/2) (n_{j} - 1/2) \ .
	\end{equation}
	The electrons at site $i$ and $j$ interact via unscreened Coulomb interaction $e^{2}/(\kappa \ r_{ij})$, where $\kappa$ is the dielectric constant and $r_{ij}$ is the distance between sites $i$ and $j$. The occupation number $n_i$ takes on values 0 or 1, corresponding to the absence or presence of an electron at site $i$, and $\phi_{i}$ is the random on-site energy.
	
	Our paper is organized as follows. In Sec. {\ref{cal}}, we present our numerical results. In Sec. {\ref{local}}, we present the results of the single-particle density of states. In Sec. {\ref{relax}}, we discuss the relaxation dynamics in the presence of unscreened Coulomb interactions and also in the case of screened Coulomb interactions.  Finally, in Sec. {\ref{discuss}}, we conclude the paper with a summary of our results.
	
	\begin{figure*}[t]
		\centering
		\includegraphics[scale=0.5]{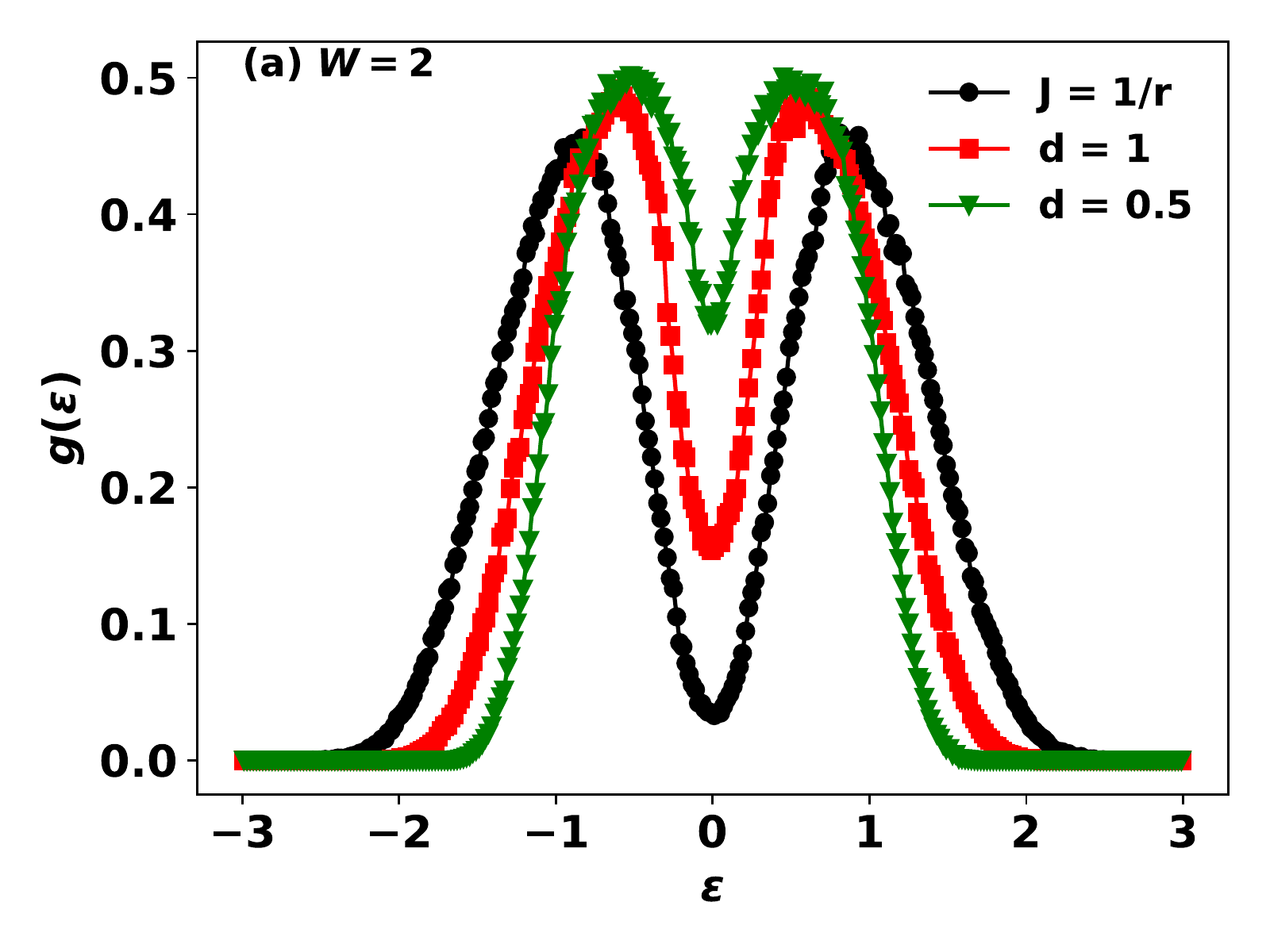}
		\includegraphics[scale=0.54]{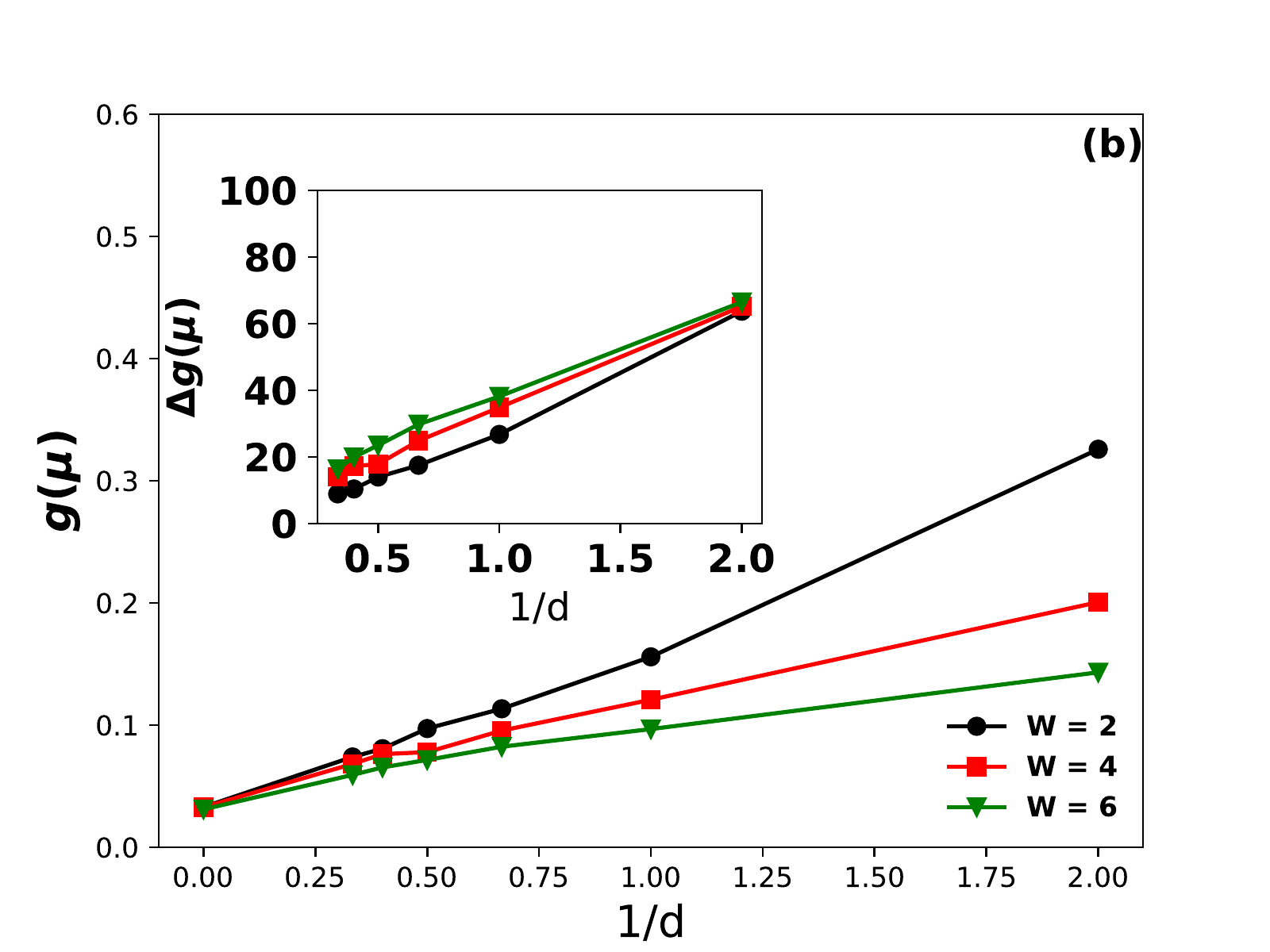}
		\caption{\label{dos_W2} (a) Histogram of the Hartree energy $\varepsilon$ (obtained using Eq.(\ref{HE})) at $W=2$ and $\beta=10$. Here $J =  1/r$ corresponds to the Coulomb interaction case as defined in Eq.(\ref{interact}), and $d = 1$ and $d = 0.5$ corresponds to the screened interaction case (see Eq.(\ref{interact})). (b) The density of states at the Fermi-level $g(\mu)$ as a function of distance $d$ (which is the separation between the metallic plate and the system). Inset shows the relative change in $g(\mu)$, which is calculated using Eq.(\ref{del_g}). }
	\end{figure*}
	
	\begin{figure*}[t]
		\centering
		\includegraphics[scale=0.5]{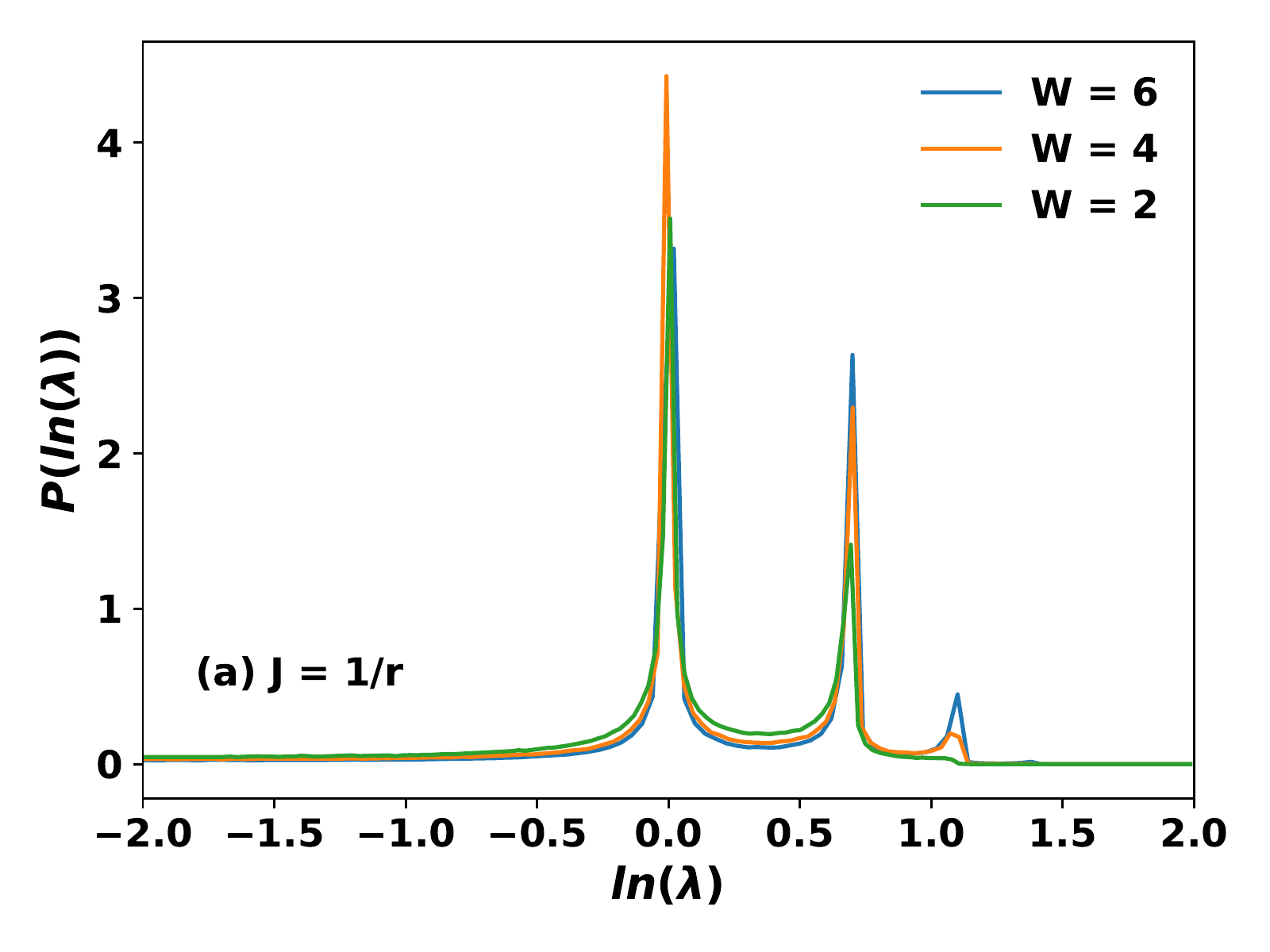}
		\includegraphics[scale=0.5]{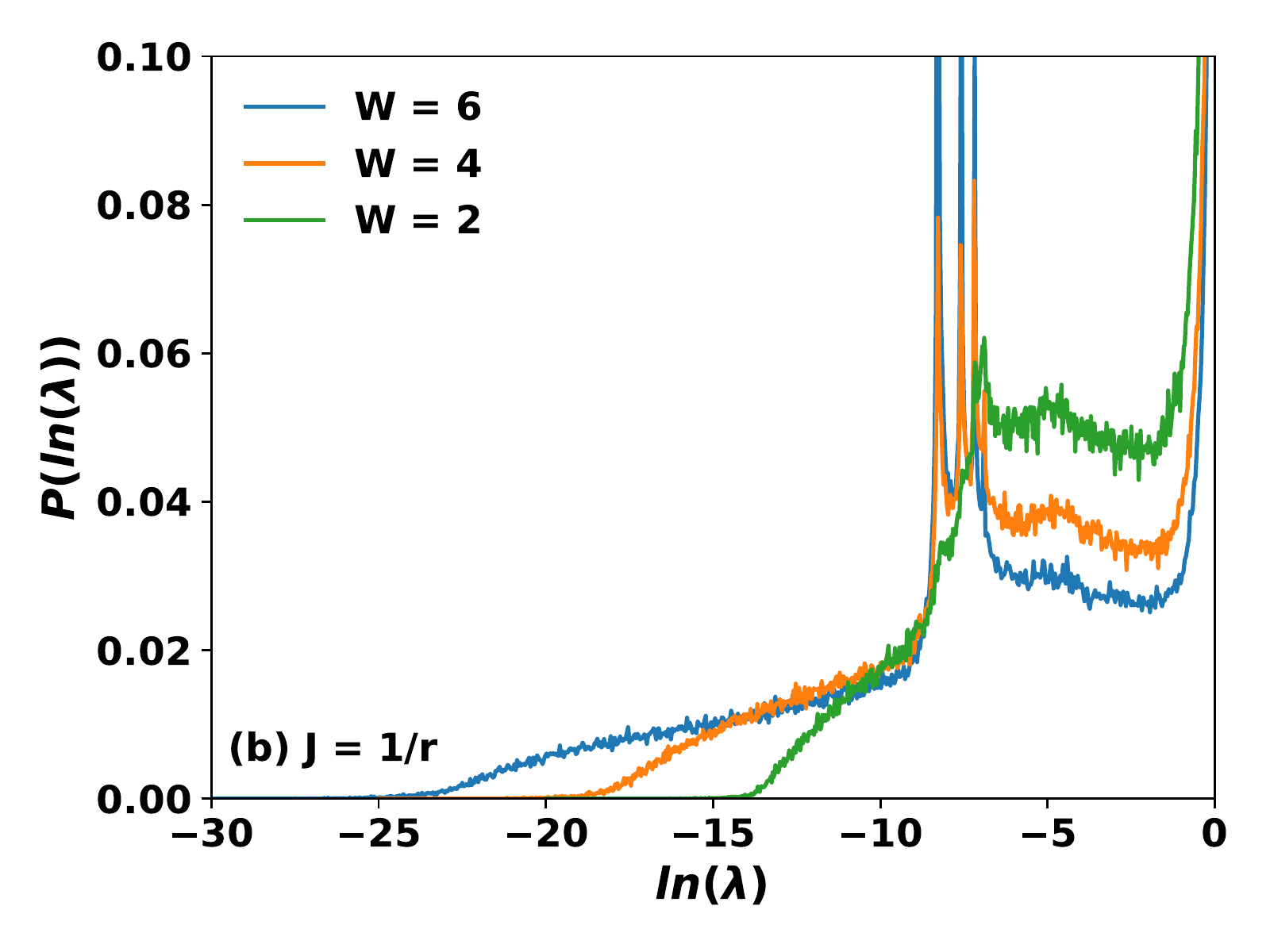}
		\caption{\label{fig:CI} Plot of the distribution of the $\ln(\lambda)$ of the dynamical matrix A obtained by solving Eq.(\ref{A-mat}) using Coulomb interactions (given in Eq.(\ref{interact})) at different disorders, (a) at short times and (b) intermediate and long times.}
	\end{figure*}
	
	\begin{figure*}
		\centering
		\includegraphics[scale=0.35]{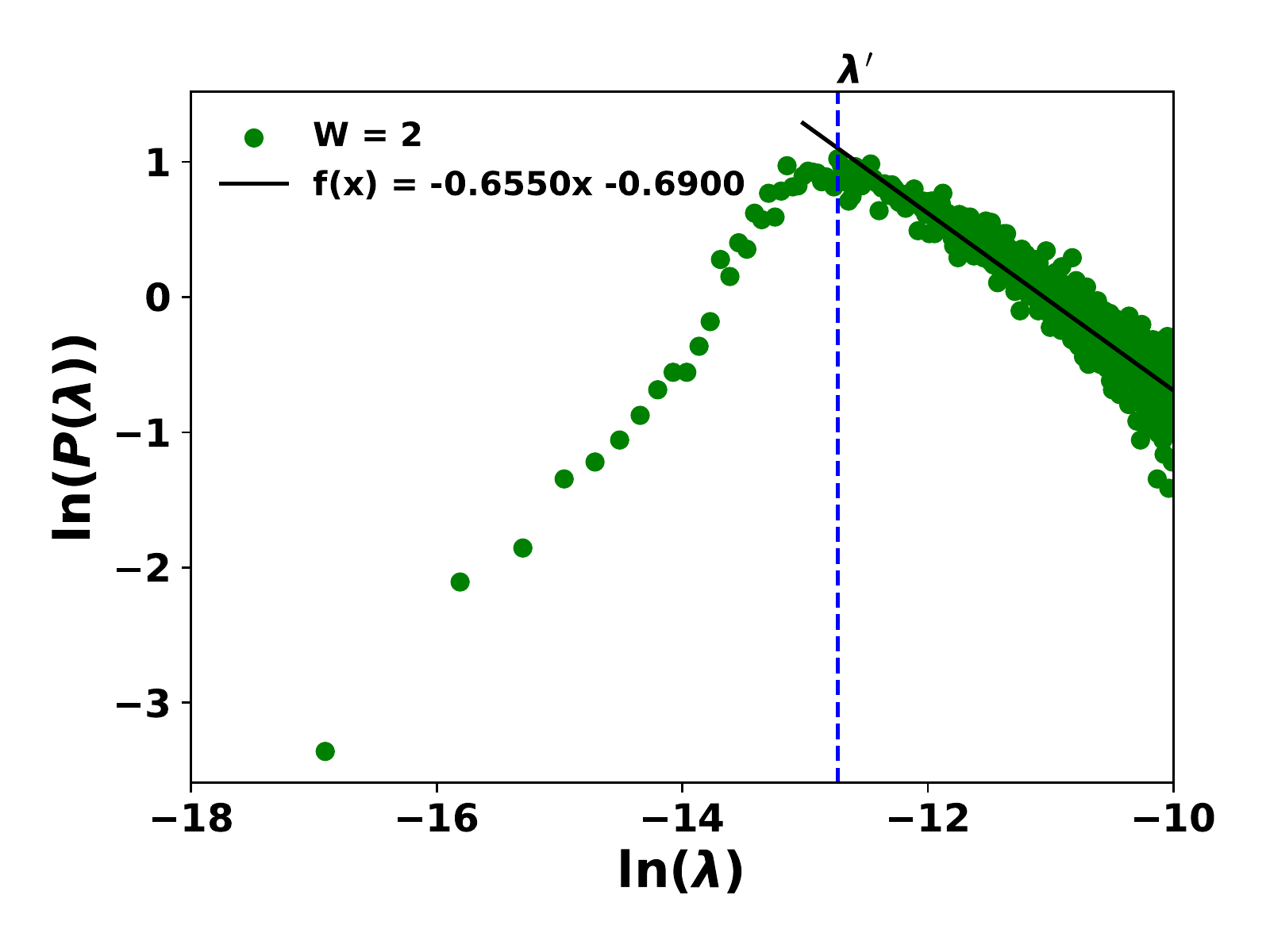}
		\includegraphics[scale=0.35]{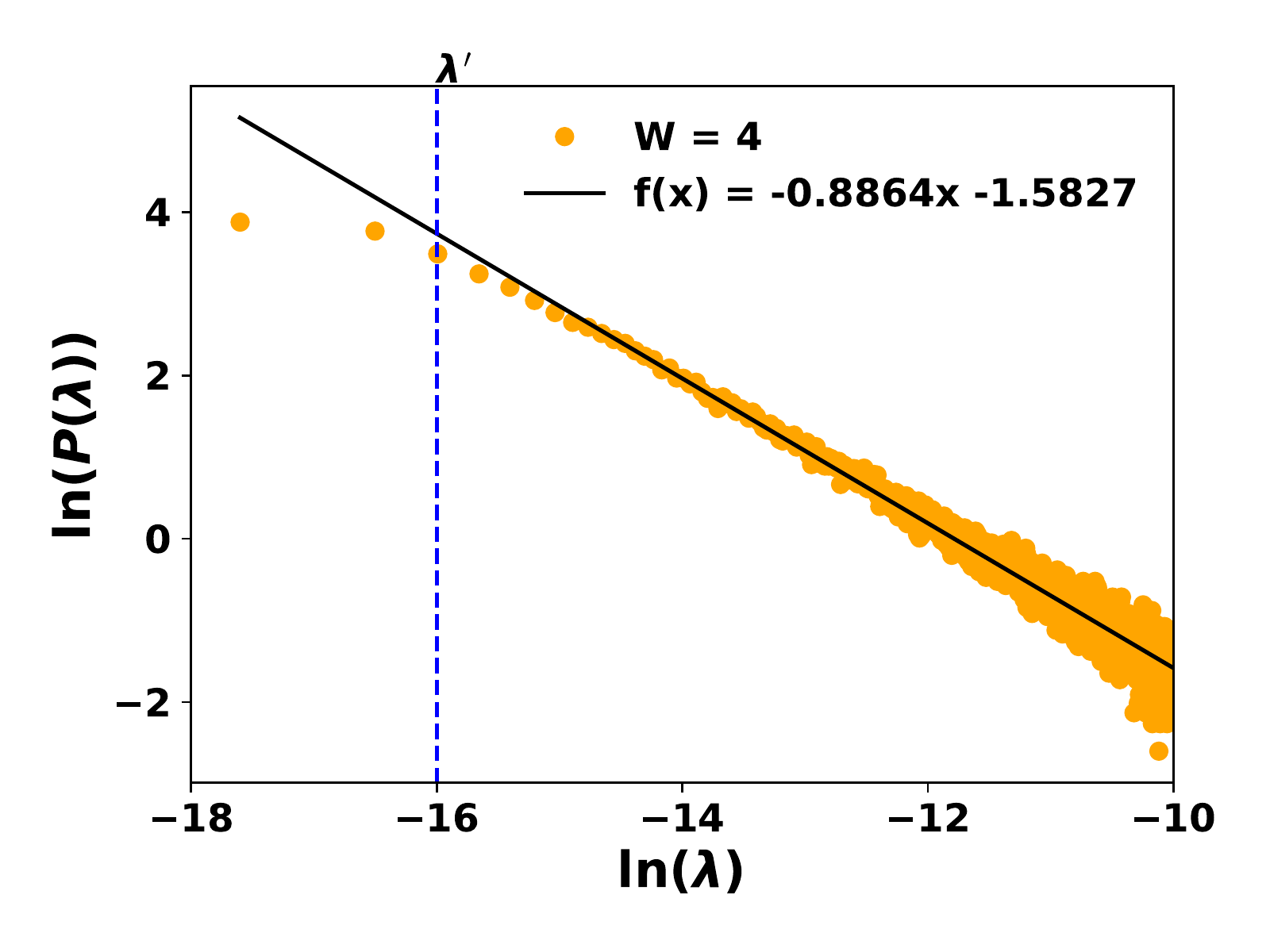}
		\includegraphics[scale=0.35]{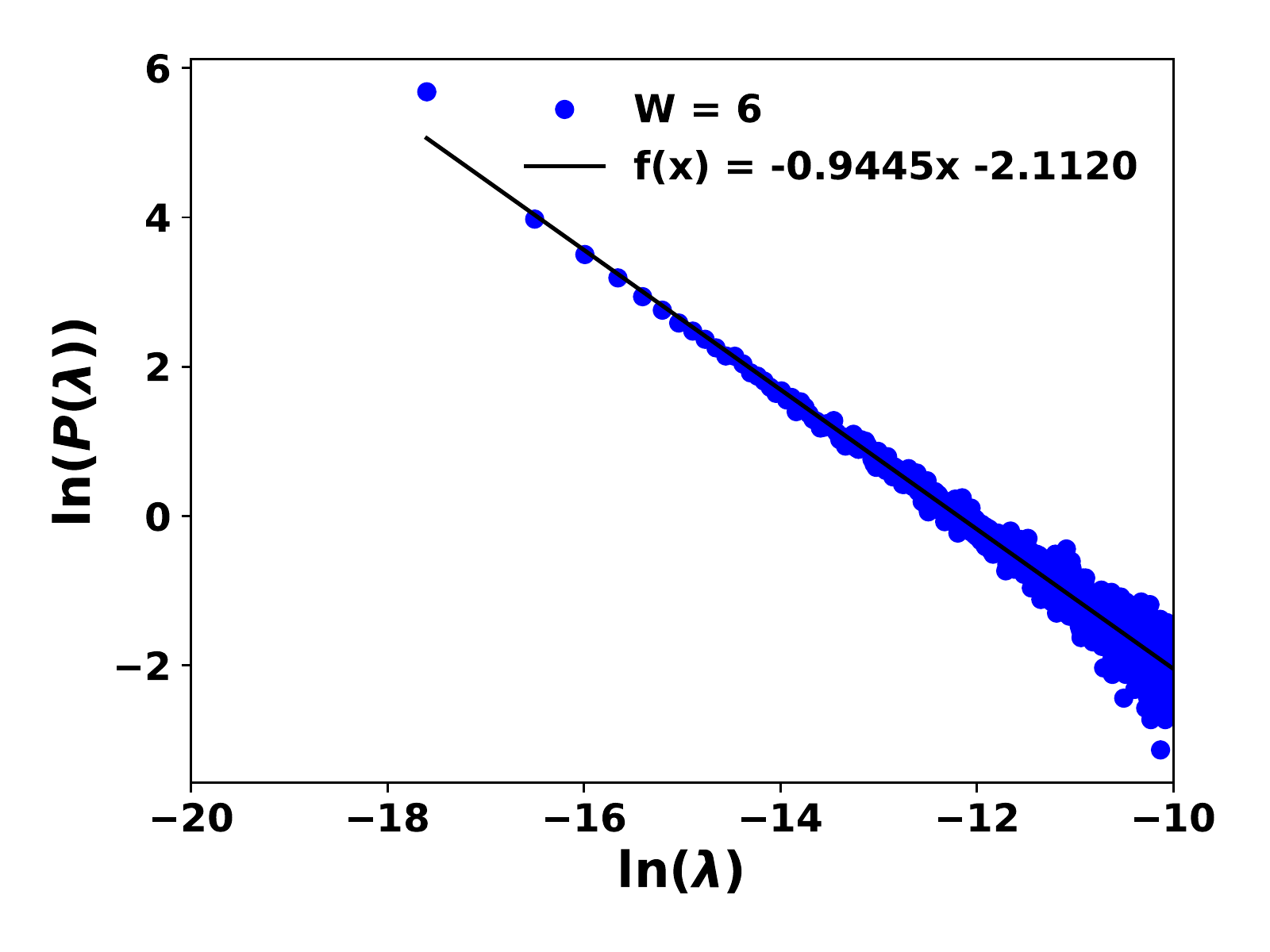}
		\caption{The relaxation law governing the behavior of small eigenvalue distribution (in Coulomb interaction case) at long times, $P(\lambda) \propto c/\lambda^{x}$ (where $x$ is provided by the slope of the straight line fit) are studied here for different disorders.}
		\label{fig:CI_fit}
	\end{figure*}
	
	\section{Results}
	\label{cal}
	\subsection{Local equilibrium}
	\label{local}
	
	The first step in calculating the relaxation dynamics is to find the local equilibrium state at a finite temperature for a given disorder realization. We use mean-field (MF) approximation to calculate the magnetization at each site $i$ at temperature $T=\beta^{-1}$. The magnetization is given by, 
	\begin{equation}
		\label{mag}
		m_{i} = tanh \, (\beta \,  \varepsilon_{i}) \ , 
	\end{equation}
	where $\varepsilon_i$ is the Hartree energy at site $i$, defined using a self-consistent equation as 
	
	\begin{equation}
		\label{HE}
		\varepsilon_{i} = \phi_{i} + \sum_{k \neq i} J_{ik} m_{i} \ .
	\end{equation}
	The interaction between sites $i$ and $k$ is given by $J_{ik}$, which can be either unscreened Coulomb interaction (CI) or screened Coulomb interaction (SI)
	
	\ba
	\label{interact}
	J_{ik} &=& \frac{1}{r_{ik}} , \quad \mbox{unscreened Coulomb interaction}, \nonumber \\
	J_{ik} &=& \frac{1}{r_{ik}} - \frac{1}{\sqrt{r^{2}_{ik} + 4d^{2}}}, \mbox{Screened interaction} \ ,
	\ea 
	where $r_{ik}$ is the distance between sites $i$ and $k$ under periodic boundary conditions, and $d$ is the separation between the metallic plate and the system (we call it the screening distance here).
	
	We solve the self-consistent equations for a square lattice with $N=1600$ sites under periodic boundary conditions. The values of $\phi_i$ are drawn from a box distribution of width $[-W/2, W/2]$, where $W$ is the disorder strength. To ensure the accuracy of our results, we average over 500 random configurations.
	
	We also investigate the influence of screening on the system's dynamics. The screened interaction takes into account the electron-electron Coulomb interaction and their images in the metal plane that is parallel to the system at a distance $d$ (screening distance). As the distance between the metallic plate and the system increases, the screening of electron-electron interaction decreases. The comparison between the density of states (DOS) with screened and unscreened interactions is shown in Fig. \ref{dos_W2}(a). In the presence of unscreened Coulomb interactions, the DOS is expected to have a soft gap around the Fermi-level at low temperatures \cite{shklovskii2013electronic,pollak2013electron}. The screening of Coulomb interactions leads to the filling of the gap.
	
	To quantify the effect of disorder and screening on the smearing of the gap, we look into the DOS at the Fermi-level ($g(\mu)$) as a function of $d$ at different disorder strengths ($W$), as shown in Fig. \ref{dos_W2}(b). We also calculate the relative change in $g(\mu)$, which is defined as	
	\begin{equation}
		\label{del_g}
		\Delta g(\mu) = \frac{g(\mu)_{SI} - g(\mu)_{CI}}{\delta_{g}} \ .
	\end{equation}
	Here, $g(\mu)_{SI}$ is the DOS at the Fermi-level due to screened interactions, $g(\mu)_{CI}$ is the DOS at the Fermi-level due to Coulomb interactions, and $\delta_g \approx W^{-1}$ is the height of the gap in the case of unscreened Coulomb interactions.
	
	\begin{figure}
		\centering
		\includegraphics[scale=0.48]{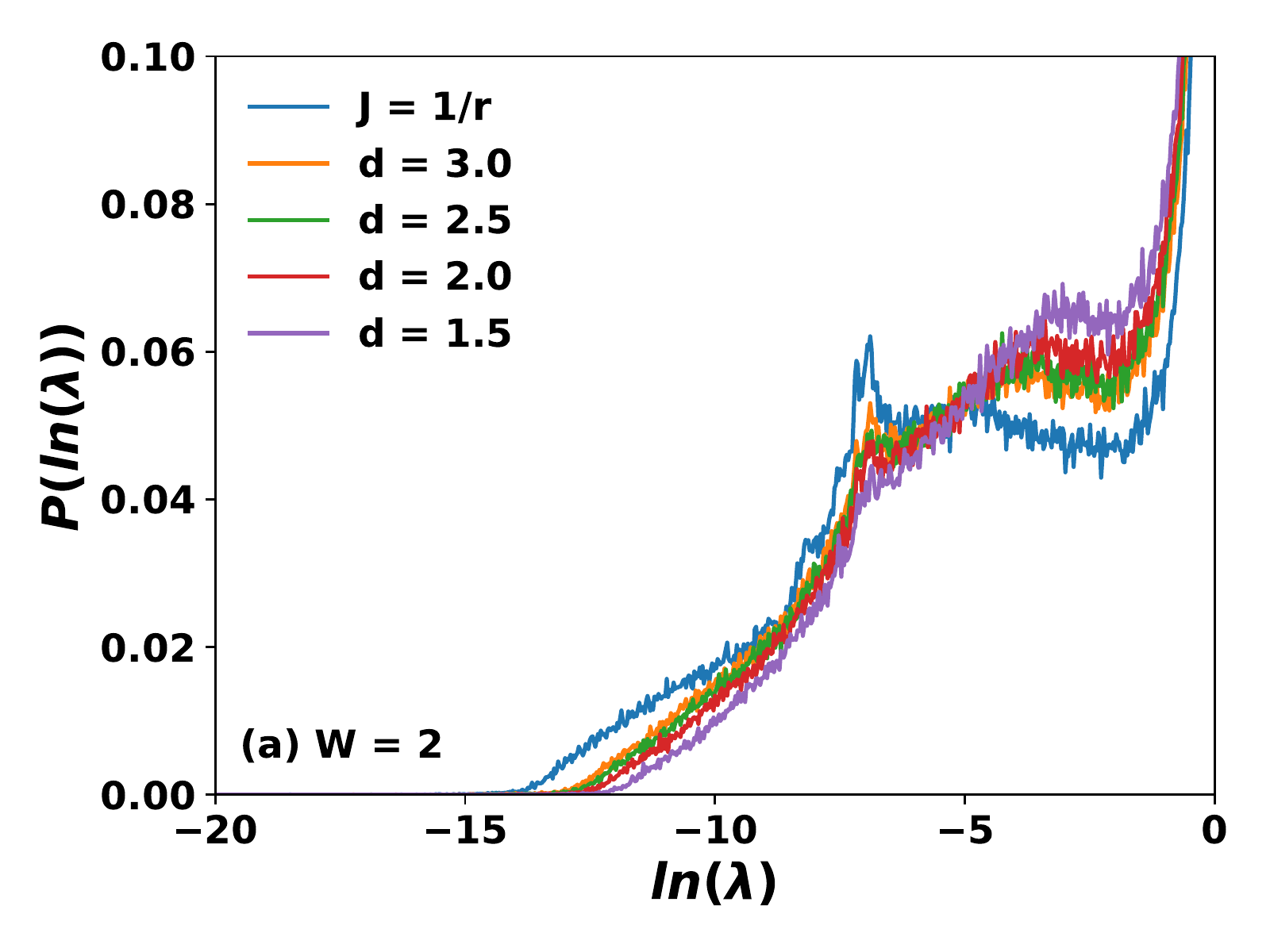}
		\includegraphics[scale=0.48]{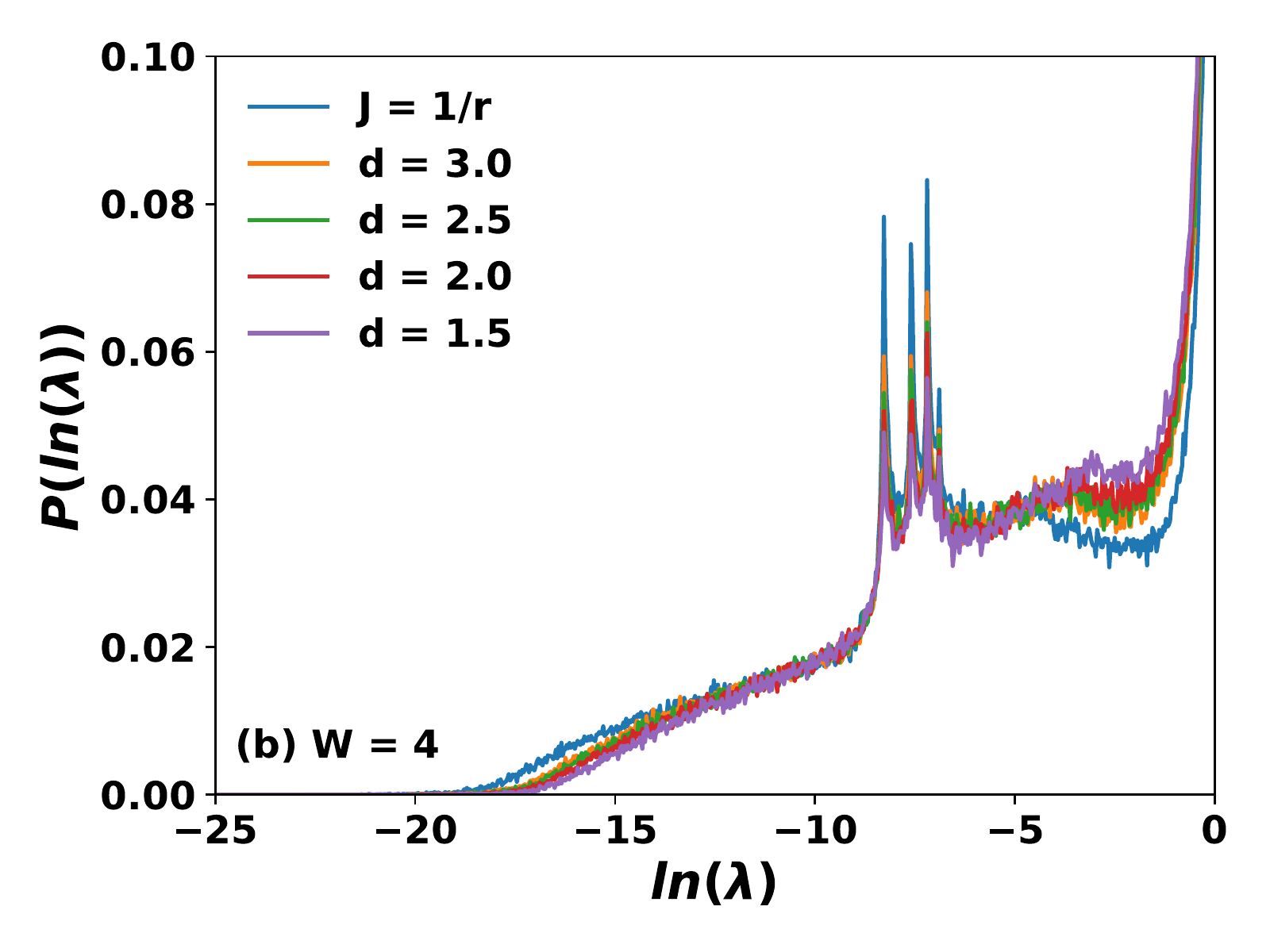}
		\includegraphics[scale=0.48]{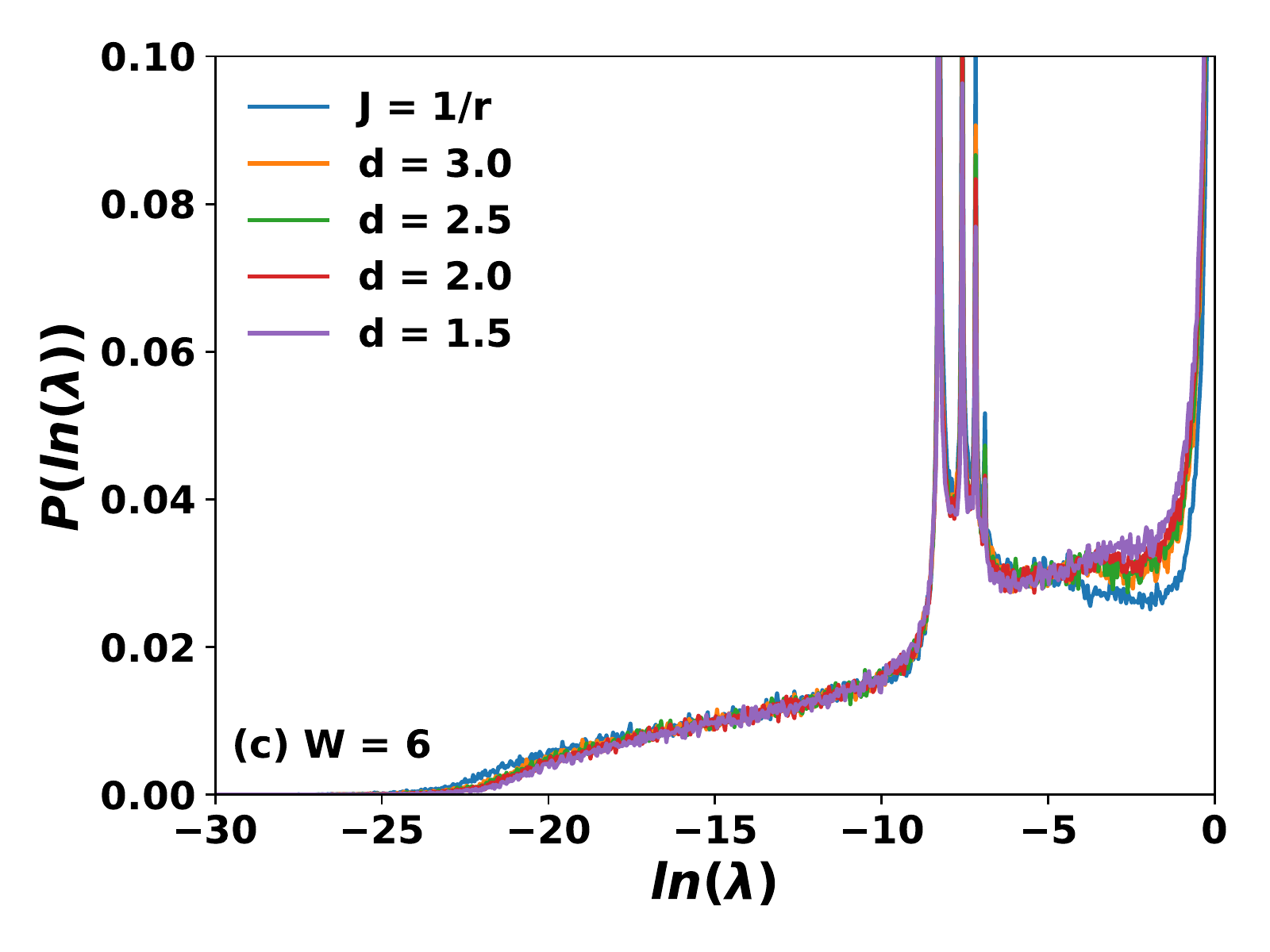}
		\caption{Log-log plot of the distribution of the eigenvalues of the dynamical matrix A obtained by solving Eq.(\ref{A-mat}) using Coulomb interactions and screened interactions. (a)-(c) Represent the intermediate and late time behavior at $W = 2,4$ and $6$ respectively. Here $J =  1/r$ corresponds to the Coulomb interaction case as defined in Eq.(\ref{interact}) and $d = 1.5$ to $d = 3$ corresponds to the screened interaction case (see Eq.(\ref{interact}))}
		\label{fig:CI_d}
	\end{figure}
	
	\begin{figure}[t]
		\centering
		\includegraphics[scale=0.5]{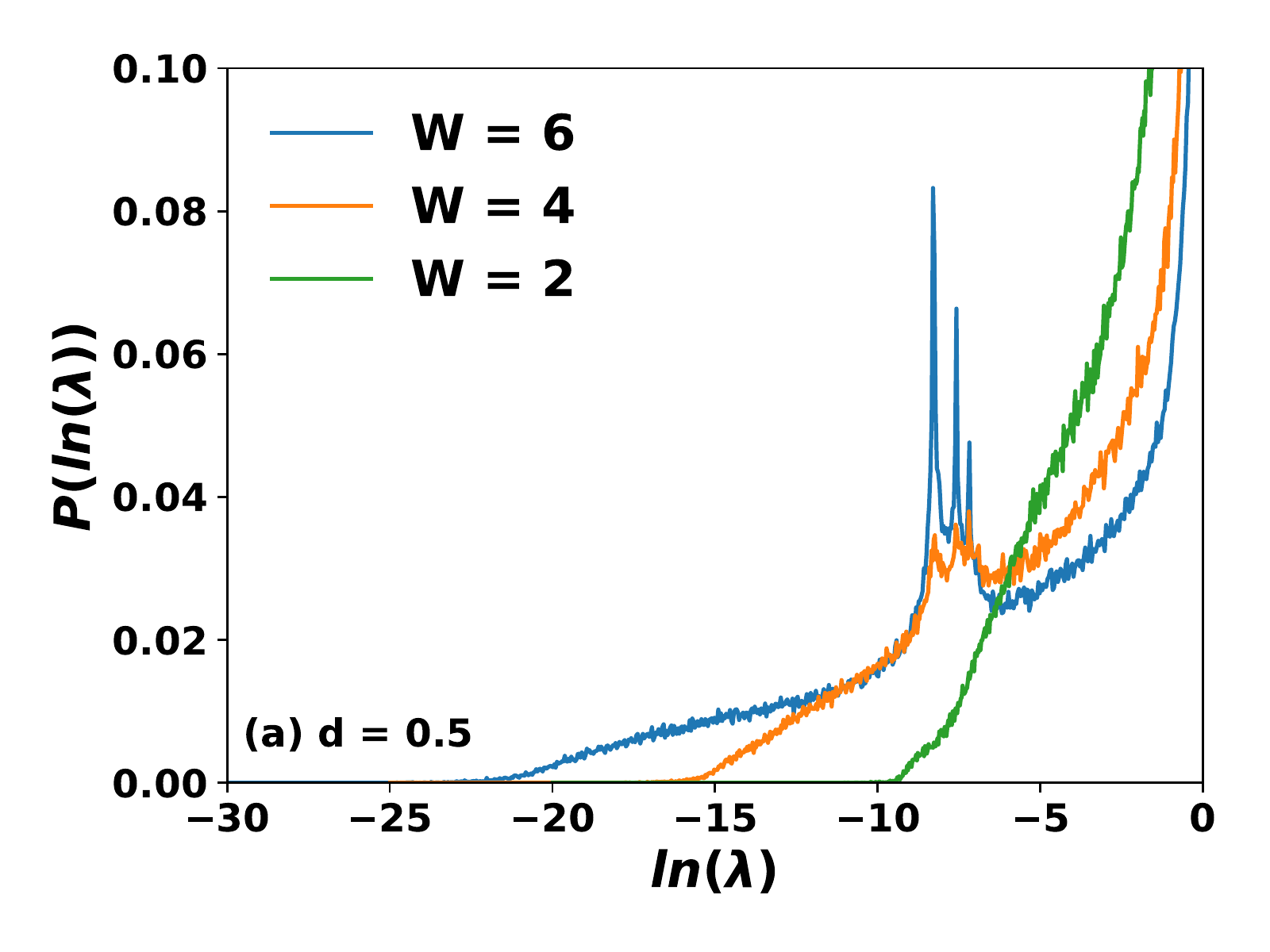}
		\includegraphics[scale=0.5]{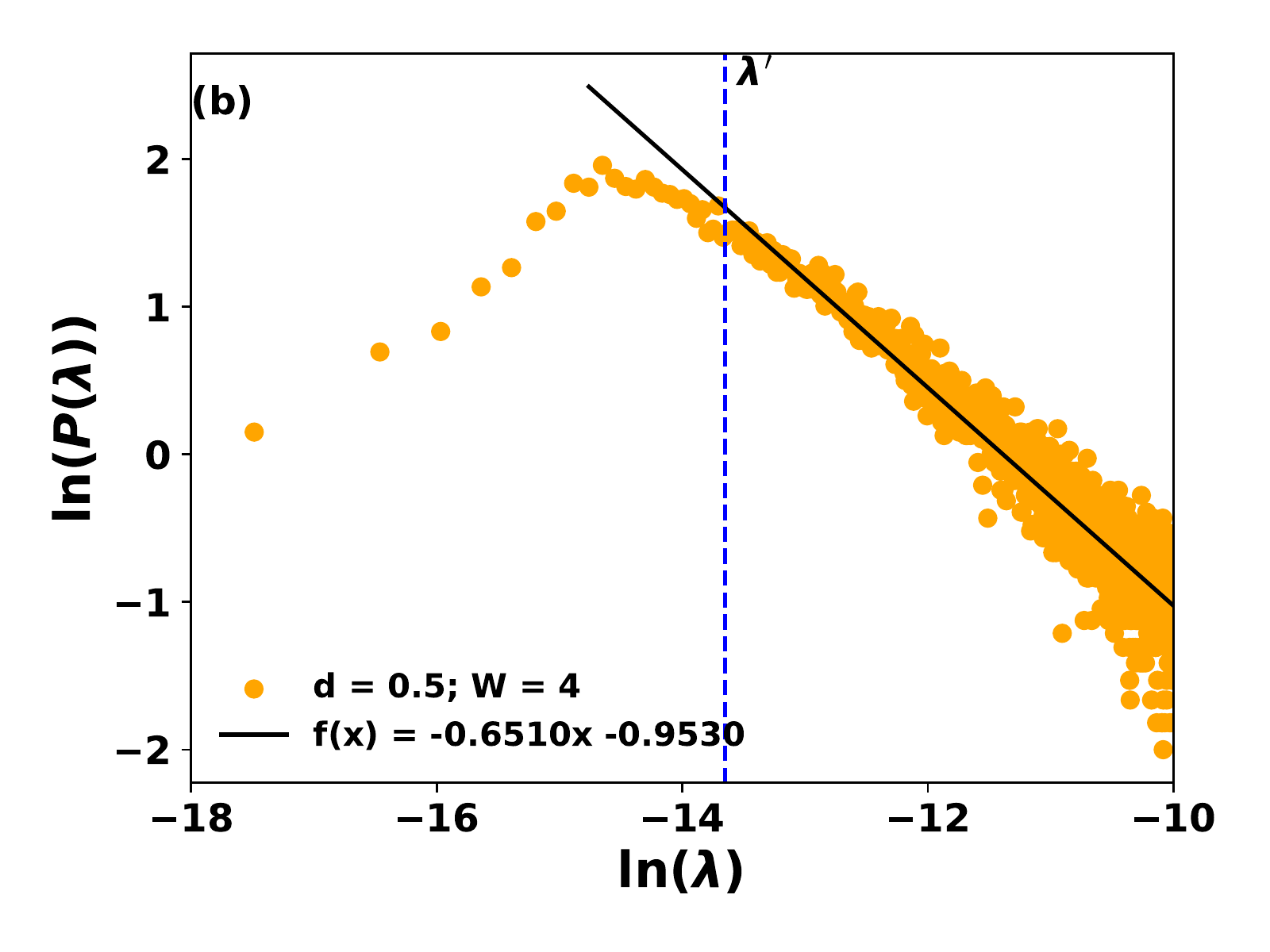}
		\includegraphics[scale=0.5]{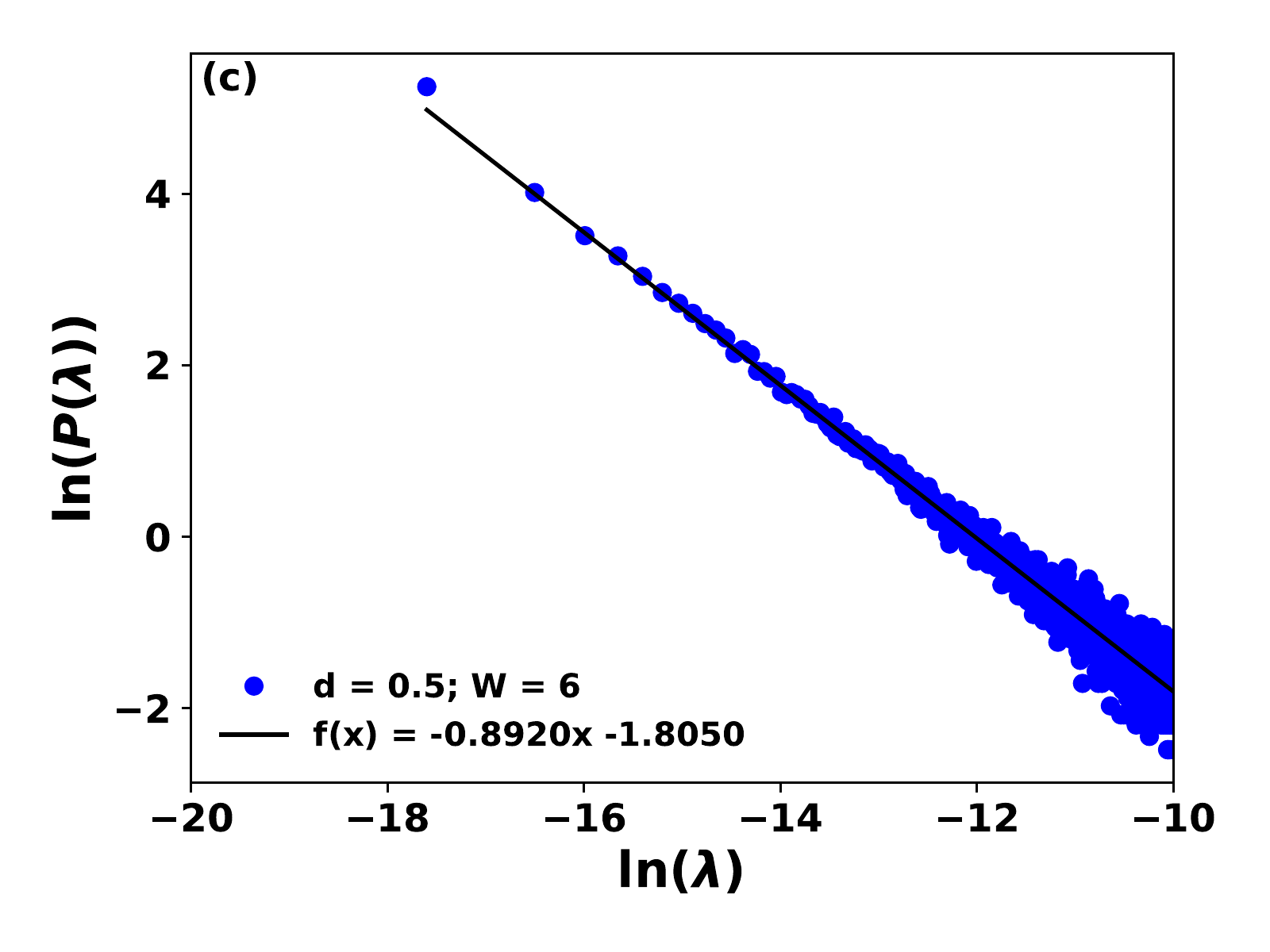}
		\caption{\label{d05} Log-log plot of the distribution of the eigenvalues of the dynamical matrix A obtained by solving Eq.(\ref{A-mat}) using screened interactions with $d = 0.5$ in Eq.(\ref{interact}).}
	\end{figure}
	
	\subsection{Relaxation dynamics}
	\label{relax}
	The conventional explanation for the sluggish dynamics in glasses is that the system struggles to overcome potential barriers and gets trapped in metastable states. This approach focuses on the energy landscape with multiple valleys \cite{pollak2006model,pollak2013electron}. In contrast, the present paper considers a single valley scenario where the system is only slightly off its local equilibrium \cite{amir2008mean,asban2017effect}. The relaxation dynamics of the system back to its local equilibrium state are then observed using a generalized master equation \cite{puri2009kinetics}.
	\ba
	\label{eq1}
	\dfrac{d}{dt} P (\lbrace n_{\mu}\rbrace, t) = - \sum_{\mu \neq \nu}  \hspace*{2mm} W_{\mu \rightarrow \nu} \hspace*{2mm} P(\lbrace n_{\mu}\rbrace,t) \nonumber \\
	+ \sum_{\nu \neq \mu}  \hspace*{2mm} W_{\nu \rightarrow \mu} \hspace*{2mm} P(\lbrace n_{\nu}\rbrace,t) \ , 
	\ea
	where $W_{\mu \rightarrow \nu}$ signifies the transition rates from state $\mu$ to $\nu$, and $P (\lbrace n_{\mu}\rbrace, t)$ is the likelihood that the system will be in state $\mu$ at the time $t$. Single-electron transfer or multiple-electron transfer can be used to describe the transition rates which conserve the particle (electron) number. In this paper, we study the evolution  by considering only the single electron transitions.
	
	When using the MF approximation, close to local equilibrium, the average occupation at site $i$ can be given as
	\begin{equation}
		\label{eq15}
		N_{i}(t) = f_{i} + \delta N_{i} \ ,
	\end{equation}
	where $f_{i}$ is the occupancy at local stable point, $f_{i} = m_{i} - 1/2$ and $\delta N_{i}$ is the deviation of average occupancy from $f_{i}$. The time evolution of the fluctuation is controlled by the matrix equation
	\begin{equation}
		\dfrac{d}{dt} \delta N_{i}= -\sum_{l} A_{i l} \hspace*{2mm} \delta N_{l} \ ,
	\end{equation} 
	where we define
	\begin{subequations}
		\label{A-mat}
		\begin{equation}
			\Gamma_{ik} = \frac{1}{2 \tau} \gamma(r_{i k}) \hspace*{2mm} f_{i} (1-f_{k}) \hspace*{2mm} f_{FD}(E^{e}_{k} - E^{e}_{i}) \ ,
		\end{equation}  
		\begin{equation}
			\Gamma_{ k i} = \frac{1}{2 \tau} \gamma(r_{k i}) \hspace*{2mm} f_{k} (1-f_{i}) \hspace*{2mm} f_{FD}(E^{e}_{i} - E^{e}_{k}) \ ,
		\end{equation}
		\begin{equation}
			A_{i i}  = \sum_{k \neq i}\, \frac{\Gamma_{i k}}{f_{i}(1-f_{i})} \ , 
		\end{equation}
		\begin{equation}
			A_{i l} = -\frac{\Gamma_{l i}}{f_{l}(1-f_{l})} - \frac{1}{T} \sum_{k(\neq l\neq i)}\, \Gamma_{i k} \hspace*{2mm} (J_{k l}-J_{i l}) \ .
		\end{equation}
	\end{subequations}  
	Here, $\gamma(r_{i k}) = \gamma_{0} e^{-r_{ik}}/\xi$, where $\gamma_{0}$ is a constant and $\xi$ is the localization length. $f_{FD}(E) = 1/(exp[\beta E] + 1)$ is the Fermi-Dirac distribution and $E_{k}^{e}$ represents the Hartree energy of site $k$ in equilibrium. The equilibrium electron transition rate from site $i$ to $k$ ($k$ to $i$) is given by $\Gamma_{ik}$ ($\Gamma_{ki}$) and $\Gamma_{ik} = \Gamma_{ki}$.
	
	The eigenvalue distribution $P(\lambda)$ of the ``A-matrix" controls the dynamics of the system, which was pushed marginally away from its local equilibrium state in this case. 
	
	The dynamics of the system can be categorized into four temporal zones, including the initial fast relaxation, slow relaxation at intermediate and long times, and the final decay to equilibrium. The initial fast relaxation is caused by the system relaxing through energy-lowering transitions between nearest neighbor sites. At intermediate and long times, the slow relaxation is due to energy-gaining transitions. The functional form of the eigenvalue distribution $P(\lambda)$ leads to different decay laws, and the system will eventually relax to equilibrium via exponential decay for times $t > 1/\lambda_{min}$, where $\lambda_{min}$ is the minimum eigenvalue of the "A-matrix." We will discuss this in detail in the next section for screened and unscreened Coulomb interactions.
	
	Slow dynamics is prevalent in systems having both interactions and disorder. Yet, distinguishing whether the dominant cause for slow dynamics lies in the interactions or in the disorder may be difficult experimentally. Specifically, in the electron glass, changing the electron density changes both disorder and interaction strengths, making their independent study difficult. Nevertheless, numerical methods can be used to investigate the individual roles of disorder and interaction. To achieve this, three different scenarios are considered here: the unscreened Coulomb interaction case, the screened Coulomb interaction case, and the case where the ratio of disorder and interaction is kept constant to examine the relaxation dynamics.
	
	\subsubsection{Unscreened Coulomb interactions}
	
	In the first scenario, the interaction strength is kept constant while varying the disorder strength to determine the effect of the disorder on slow relaxation. The distribution $P(ln(\lambda))$ of the eigenvalue $\lambda$ of the ``A-matrix" is displayed in Fig.(\ref{fig:CI}). The eigenvalues have been scaled by the factor $exp(-1.0/\xi)$. The plot shows that $P(ln(\lambda))$ displays peaks at high eigenvalues (initial times) corresponding to $\lambda = 1$ and $2$. These peaks represent an electron relaxing to the nearest neighbor (NN) site through an energy-lowering transition in the case of a single such available site ($\lambda=1$) and two such available sites ($\lambda=2)$. At intermediate times, in the regime $\ln(\lambda) = -2$ to $\ln(\lambda) = -5$, the distribution $P(ln(\lambda))$ is approximately parallel to the x-axis as shown in Fig.(\ref{fig:CI}(b)). This implies that $P(\lambda) \approx c/\lambda$, and thus the fluctuations in the system decay by logarithmic decay law $(\delta n(t) \sim ln(t))$ at intermediate times for all disorders.
 
	Subsequently, the relaxation process in the system occurs through electron transitions to the next nearest neighbor site, resulting in energy-lowering transitions. This phenomenon gives rise to peaks in the probability distribution at approximately $\ln{\lambda} \approx -8$ (see Fig.\ref{fig:CI}(b)). Each peak corresponds to a specific number of available transition sites: one peak represents a single possible transition, while the other two peaks correspond to two and three available transition sites. This behavior is analogous to the peaks observed in Fig.\ref{fig:CI}(a) for nearest neighbor transitions.

    As time progresses, particularly at long timescales, the system undergoes relaxation through energy-gaining transitions within the range of $e^{-\lambda_{min}} < \lambda < e^{-10}$. This region can be further divided into two parts. Initially, there is a period of slow relaxation ($\ln(\lambda^{\prime}) < \ln(\lambda) < -10$), followed by a more pronounced decrease towards $\lambda_{min}$. We observe that the system relaxes slowly until the time $t^{\prime} = 1/\lambda^{\prime}$, where the specific value of $\lambda^{\prime}$ depends on the disorder (as depicted in Fig.\ref{fig:CI_fit}). Subsequently, the system exhibits exponential decay after the time $t_r = 1/\lambda_{min}$. It is worth noting that these two distinct timescales, $t^{\prime}$ and $t_{r}$, have also been observed experimentally \cite{ovadyahu2019screening}.
	
	To determine the slow relaxation law in the late time regime, $P(\lambda)$ vs. $\ln(\lambda)$ is shown in Fig.(\ref{fig:CI_fit}). It is noted that the plot of $P(\lambda)$ against $\lambda$ on a log-log scale is a straight line when $\lambda$ is greater than $\lambda^{\prime}$, with an absolute value of the slope less than $1$. This indicates that the relaxation follows a power-law decay ($\delta n(t) \sim t^{-\alpha}$). The absolute value of the slope increases as $W$ becomes larger, and for $W = 4$ and $6$, it approaches one, indicating that the behavior is close to logarithmic decay in these cases. The crossover from logarithmic decay to power-law behavior with $\alpha = 0.2$ with time has been observed in experiments \cite{vaknin2000heuristic}, while in some experiments \cite{vaknin2002nonequilibrium}, only $\ln(t)$ behavior is observed.
	
	We also observe (see Fig.\ref{fig:CI}(b)) that $P(\lambda)$ for the intermediate times decreases as the disorder increases. This implies that as the disorder increases, the probability of an electron finding a hole to be excited to with only a small energy gain at the nearest neighbor site decreases. These electrons jump to higher energy holes (at long times) to relax as the disorder increases, leading to a decrease in the value of $\lambda_{min}$. These effects can be explained by a decrease in the number of states around the Fermi level as the disorder increases. The number of holes in an energy range $E$ above the Fermi level is proportional to $E/W$ for energies inside the Coulomb gap and $1/W$ for sites outside the gap. Thus the probability of an electron finding a hole ($\Delta E > 0$ transitions) with energy $E$ decreases as disorder increases. For energy-gaining transitions, $\lambda \propto e^{-\beta \Delta E}$, so $P(\lambda)$ and $\lambda_{min}$ decreases with an increase in disorder at intermediate times. The system will eventually relax to a local equilibrium state via the exponential law $\delta n(t) = e^{-t/\tau}$. The value of $\tau = 1/\lambda_{min}$ will increase with increasing disorder. The conundrum of whether the disorder or the Coulomb gap (due to long-range Coulomb interactions) is the primary cause of the slow relaxation arises from the fact that $J/W$ is decreasing as $W$ increases.
 
	\begin{figure*}
		\centering
		\includegraphics[width=0.45\textwidth]{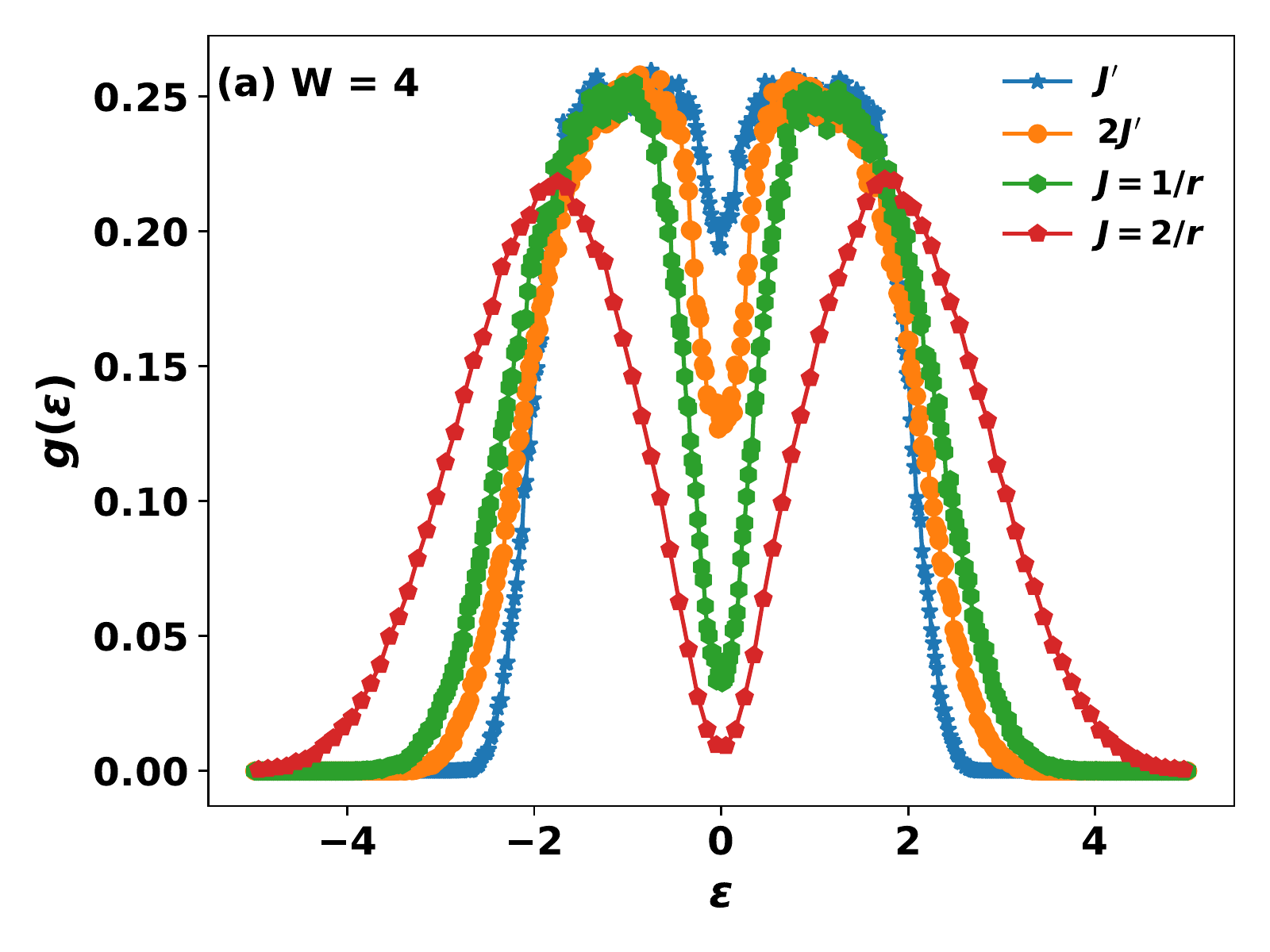}
		\includegraphics[width=0.45\textwidth]{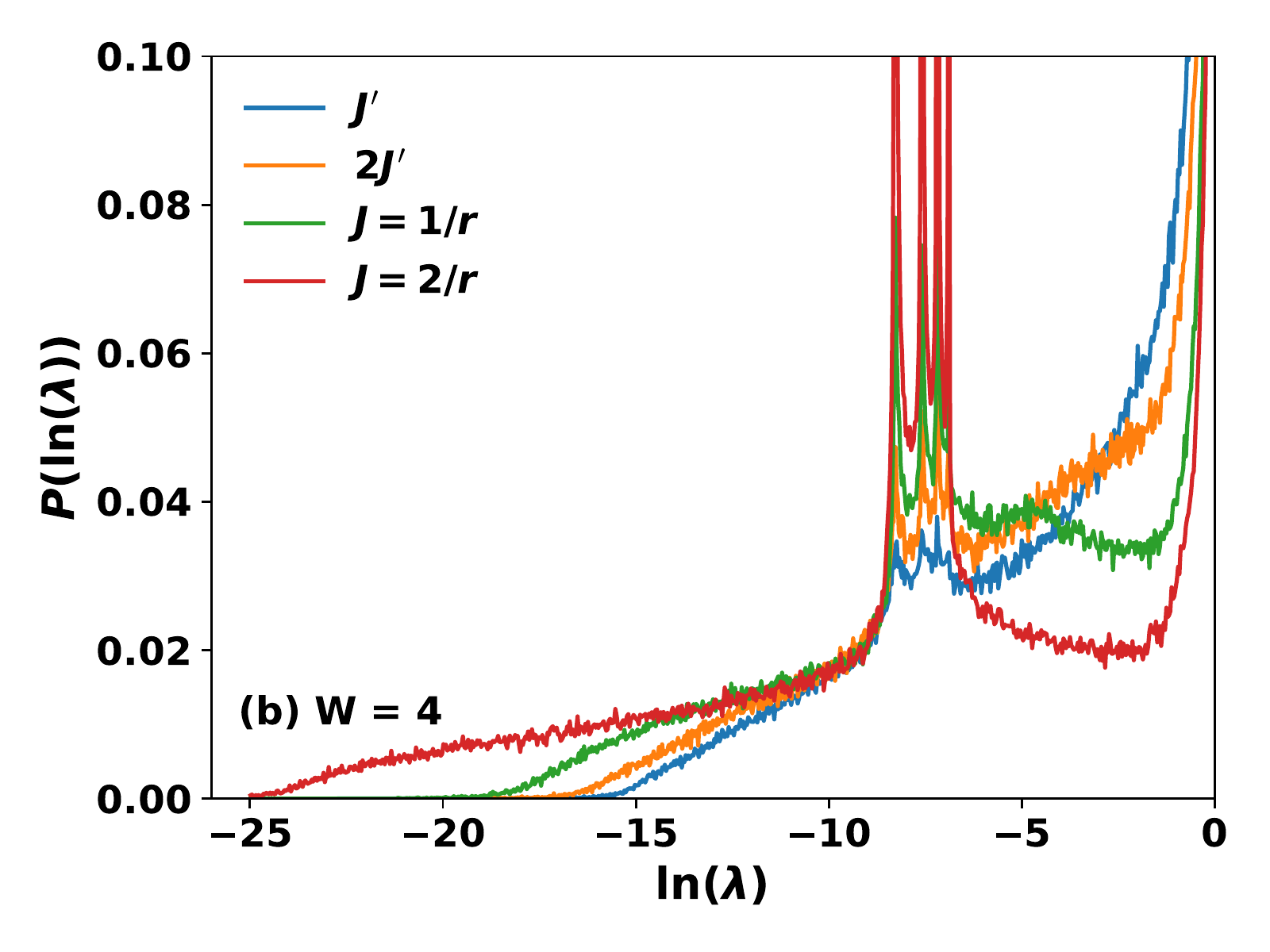}		
		\caption{(a) Histogram of the Hartree energy $\varepsilon$ (obtained using Eq.(\ref{HE})) for different interaction strengths at $W = 4$. (b) Log-log plot of the distribution of the eigenvalues of the dynamical matrix $A$ obtained by solving Eq.(\ref{A-mat}). $J^{\prime} = 1/r_{ik} - 1/\sqrt{r^{2}_{ik} + 4d^{2}}$ and $2J^{\prime} = 2/r_{ik} - 2/\sqrt{r^{2}_{ik} + 4d^{2}}$ corresponds to the screened interaction case where $d = 0.5$. $J = 1/r_{ik}$ and $J = 2/r_{ik}$ corresponds to the Coulomb interaction case.}
		\label{SI_dos_compare} 
	\end{figure*} 
	
	\begin{figure*}
		\centering
		\includegraphics[scale=0.38]{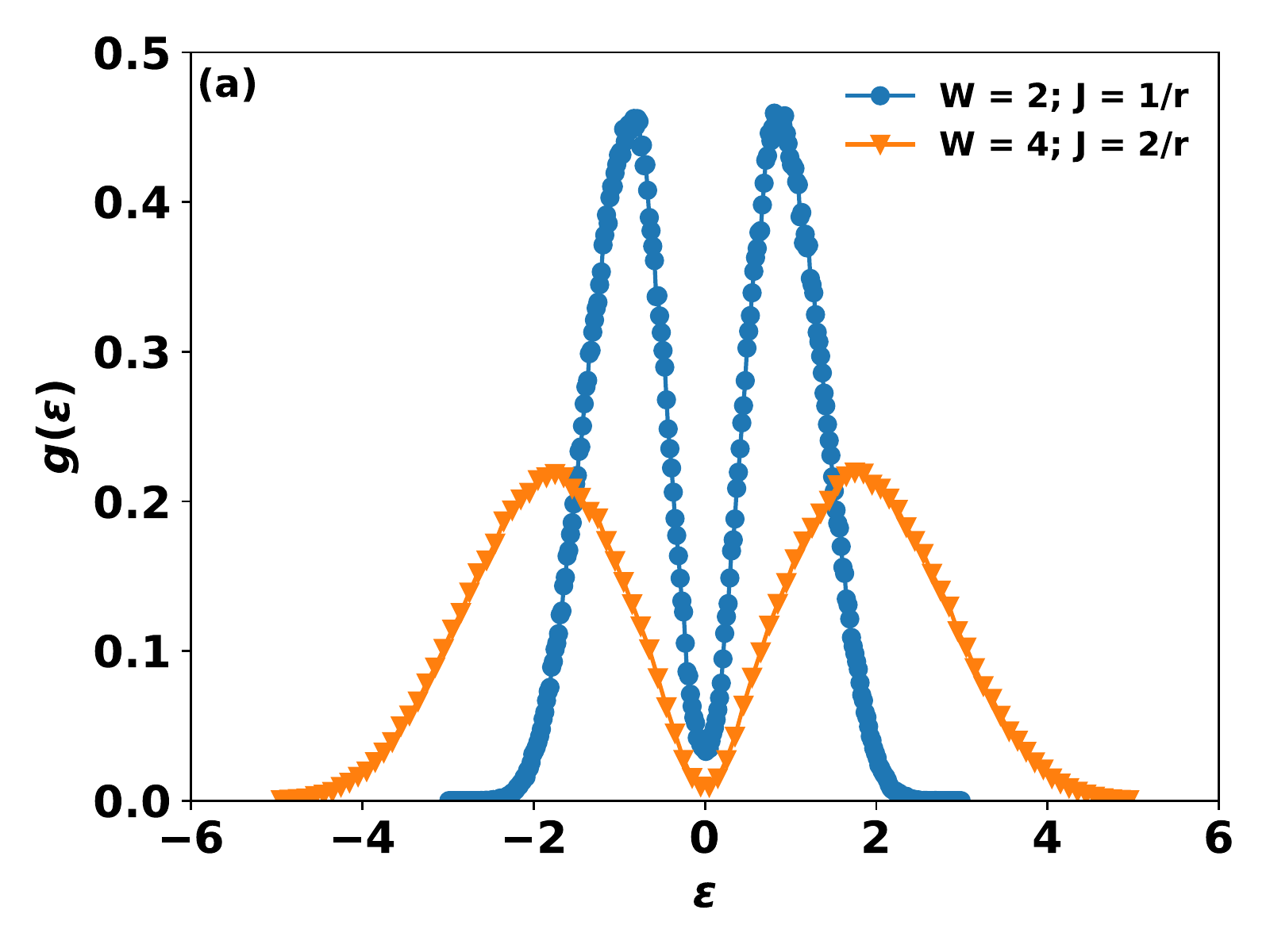}
		\includegraphics[scale=0.38]{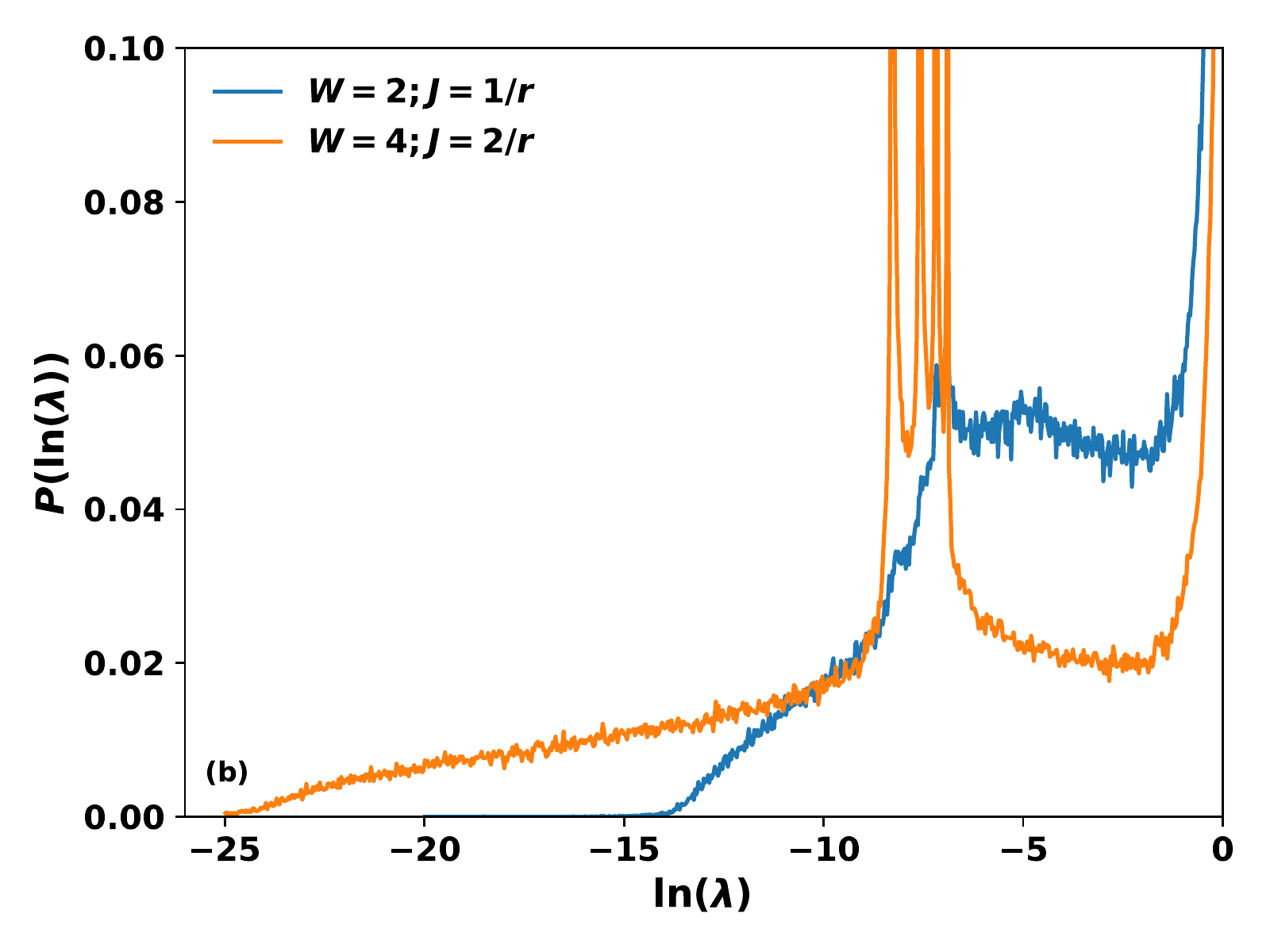}
		\includegraphics[scale=0.38]{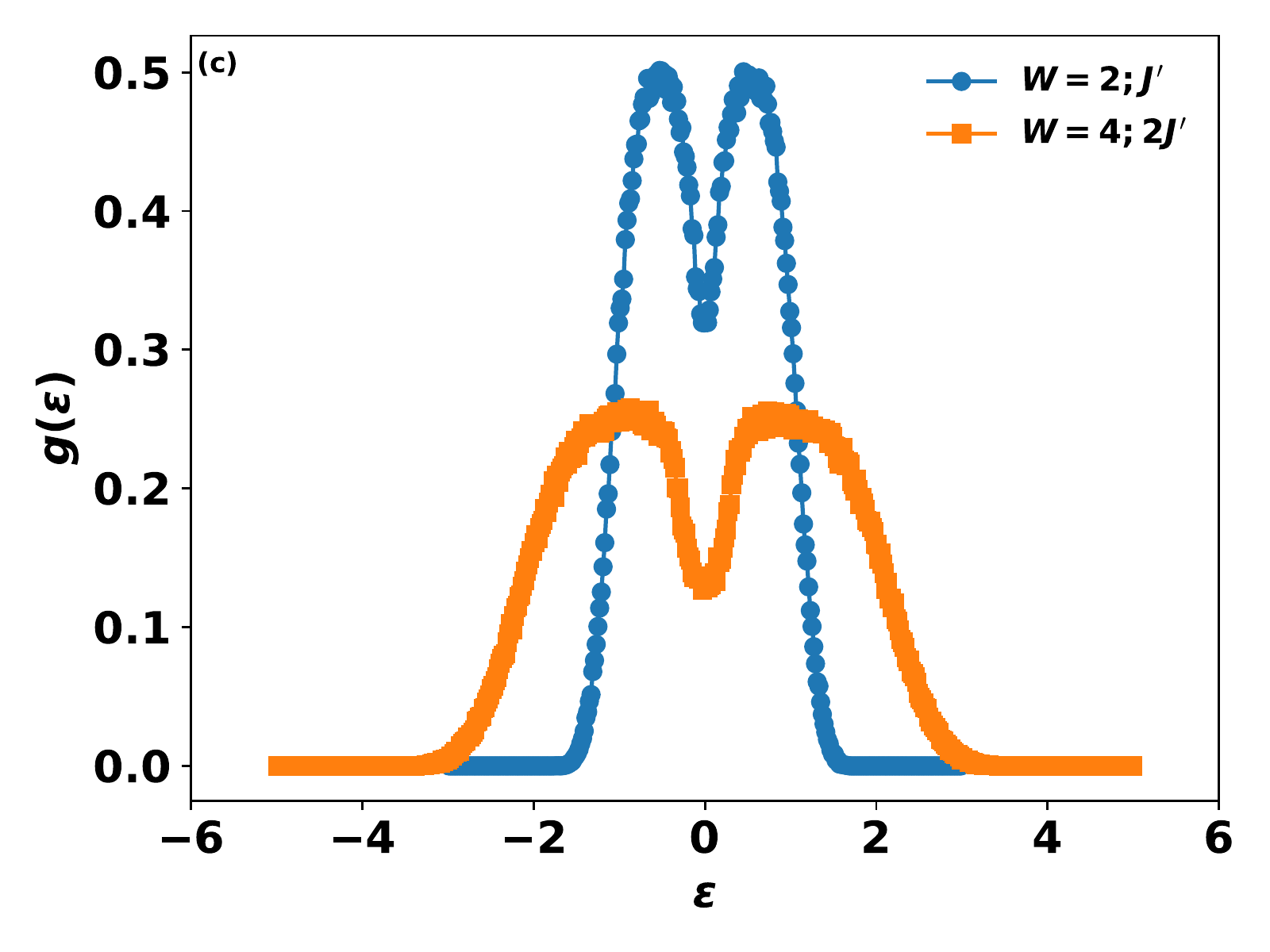}
		\includegraphics[scale=0.38]{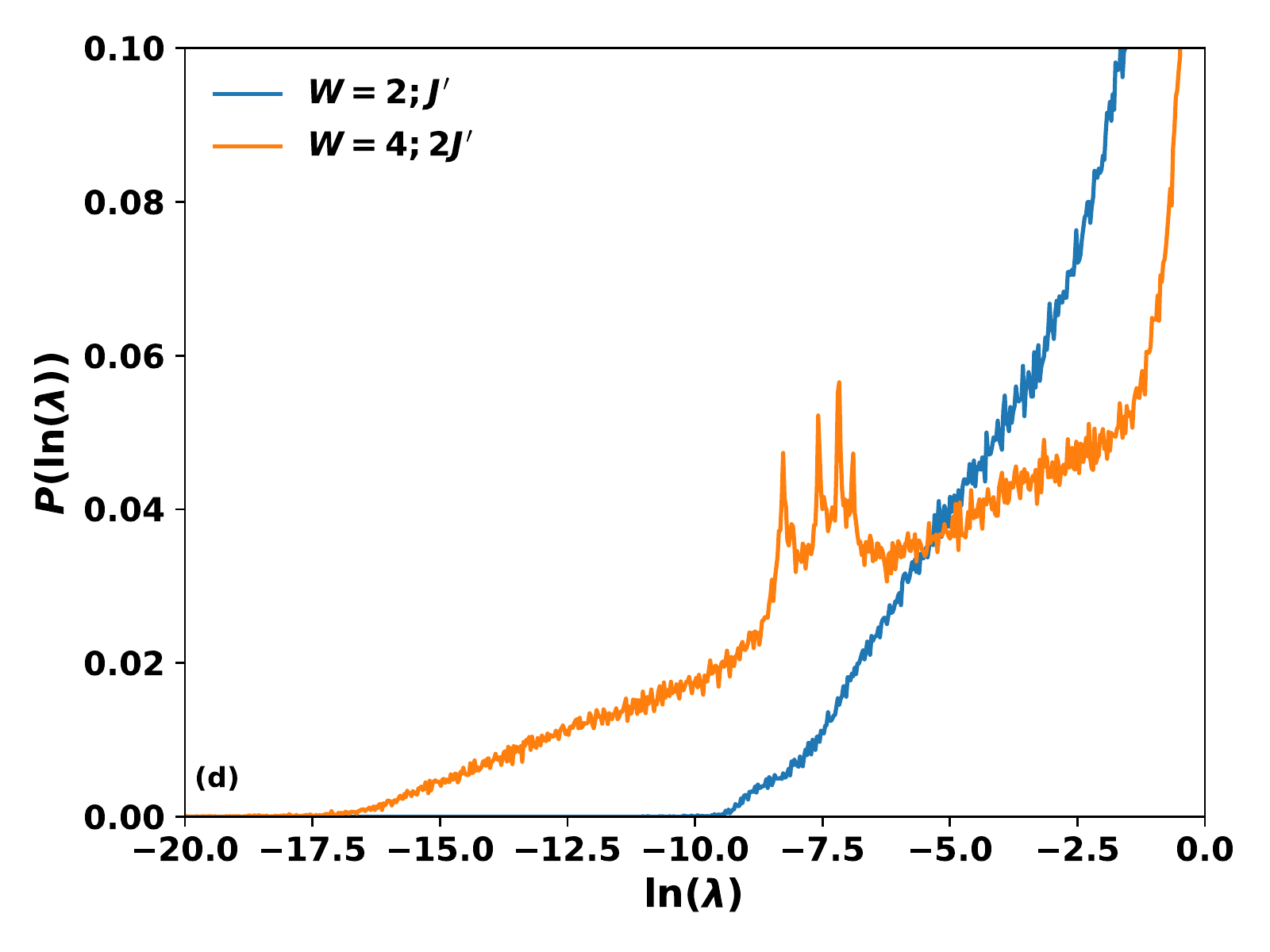}
		
		\caption{(a)-(c) Histogram of the Hartree energy $\varepsilon$ (obtained using Eq.(\ref{HE})) for different disorder values and interaction strengths. (b)-(d) Log-log plot of the distribution of the eigenvalues of the dynamical matrix $A$ obtained by solving Eq.(\ref{A-mat}). $J^{\prime} = 1/r_{ik} - 1/\sqrt{r^{2}_{ik} + 4d^{2}}$  and $2J^{\prime} = 2/r_{ik} - 2/\sqrt{r^{2}_{ik} + 4d^{2}}$ corresponds to the screened interaction case where $d = 0.5$. $J = 1/r_{ik}$ and $J = 2/r_{ik}$ corresponds to the Coulomb interaction case.} 
		\label{cont_ratio_compare} 
	\end{figure*} 
	
	\subsubsection{Screened Coulomb interactions}
	
	In order to separate the effects of disorder and interaction, we now consider the case where the disorder is constant, but the interactions change as a result of the addition of a screening plate. The effect of disorder strength ($W$) and screening length ($d$) on the dynamics of the system is analyzed by keeping the disorder strength constant while screening the Coulomb interactions using the screened interaction specified in Eq.(\ref{interact}). The behavior of $P(\lambda)$ is studied for various screening lengths and disorders, and the results are illustrated in Fig.(\ref{fig:CI_d}).
	
	At a fixed disorder level, it has been observed that the slope of $P(\ln(\lambda))$ versus $\ln(\lambda)$ increases as the screening parameter $d$ decreases, specifically for intermediate and large time scales. Furthermore, the value of $\lambda_{min}$, which represents the inverse of the longest relaxation time, also increases as $d$ decreases. These findings suggest that the relaxation process becomes faster as the screening of Coulomb interaction increases. It is worth noting that the changes in the slope value and $\lambda_{min}$ are relatively small when the disorder is sufficiently high, as depicted in Fig.4(c).
	
	The effect of screening on the relaxation dynamics is attributed to the smearing of the gap in the density of states (DOS) due to screening, as seen in Fig.(\ref{dos_W2}(a)). The screening leads to a sufficient number of electrons (holes) close to the Fermi level, allowing the system to relax to a low-energy state more effectively. The filling of the gap is found to be around $10-25\%$ for screening lengths corresponding to $d \geq 1.5$, as shown in Fig.(\ref{dos_W2}(b)). For these values of $d$, the difference in the slopes between screened and unscreened cases at $W = 4$ and $6$ is small, and hence the system will decay via a nearly logarithmic law, similar to experimental observations \cite{ovadyahu2019screening} where Coulomb interactions were screened by a metallic plate, resulting in a $12-23\%$ filling of the Coulomb gap.
	
	The effect of very strong screening ($d = 0.5$) on the dynamics is also studied. In Fig.\ref{dos_W2}(b), one observes that at $d = 0.5$, the gap is about $60\%$ filled for all disorders. Comparing (see Fig.\ref{d05}(a) and Fig.\ref{fig:CI_d}, one observes that $\lambda_{min}$ increases substantially by a factor of approximately $e^{3}$ with respect to the unscreened case. This implies that the system will show slow relaxation for a much shorter period of time, which may not be observable experimentally. For the smallest disorder considered ($W = 2$), the system will show no logarithmic decay, and for $W = 4$, the system will decay via power law ($\delta n(t) = 1/t^{\alpha}$) for intermediate and long times with $\alpha$ values of 0.1136 and 0.3490, respectively. At high disorder ($W = 6$), the system will decay via a nearly logarithmic law (with $\alpha$ values of 0.0555 and 0.1080 at intermediate and long times, respectively), showing that slow relaxation can happen without strong interaction effects consistent with the experimental findings \cite{ovadyahu2019screening}.
	
	We also analyzed the effect of doubling the interaction strength while keeping the disorder same in both screened and unscreened cases. Here, we study a $W = 4$ case (see Fig.(\ref{SI_dos_compare})) in which the unscreened and strongly screened ($d = 0.5$ case) Coulomb interaction exhibits logarithmic and power-law decay respectively at intermediate times. For the unscreened case, the relaxation time $\tau$ increases by a factor of approximately $e^{7}$ on doubling the interaction strength. This slowdown in the relaxation process is due to the doubling of the width of the Coulomb gap leading to a significant decrease in DOS around the Fermi level. This behavior has also been observed in experiments where the strength of interaction was increased, keeping the disorder strength constant. \cite{ovadyahu2019screening}. At later times, the power law exponent $\alpha$ decreases significantly, and the decay becomes approximately logarithmic. Comparison of $2J^{\prime} = 2/r - 2/\sqrt{r^{2} + 4d^{2}}$, $W = 4$ with $J^{\prime} = 1/r - 1/\sqrt{r^{2} + 4d^{2}}$, $W = 4$ case where $d = 0.5$ (see Fig.(\ref{SI_dos_compare}(b))) shows that relaxation in both the cases is similar. Thus in the case of strong screening, the interaction is now short-range, and the degree of relaxation is mainly determined by the degree of disorder. Our results suggest that doubling the interaction without screening has a far more substantial effect on the dynamics than doubling the interaction in the strongly screened scenario. Another observation that is made from Fig.\ref{SI_dos_compare}(a) is that the width of the Coulomb gap is roughly comparable for the two screened cases and the $J = 1/r$ case. The only notable variation is the Coulomb gap's dip. While the gap is well-formed in the unscreened case, it is substantially filled in the two screened cases. As a result, the variation in $\tau$ values for these three scenarios is minimal, but logarithmic decay at intermediate times can only be seen for the unscreened Coulomb interaction. We provide a rough explanation for this behavior. The width of the Coulomb gap and the single-particle density of states affects the distribution's lowest eigenvalue. On the other hand, the depletion of the density of states (DOS) at the Fermi level is responsible for the logarithmic time dependence observed at intermediate times for the unscreened Coulomb interaction.
	
	\subsubsection{Constant J/W ratio}
	
	Finally, in the third scenario, both the disorder and interaction strength are increased while keeping their ratio constant. This scenario is expected to be similar to the experimental situation where the carrier concentration and disorder in a sample increase. 
	
	The results show that when both the disorder and interaction strength are doubled for the unscreened interaction case, the relaxation dramatically slows down. This is evident from the increase in $\tau$, which is roughly $e^{10}$ times greater when J = 2/r and W = 4 compared to when J = 1/r and W = 2 (as shown in Fig.\ref{cont_ratio_compare}(b)). The decrease in DOS around the Fermi level, as shown in Fig.\ref{cont_ratio_compare}(a), is attributed to both an increase in disorder and the widening of the Coulomb gap. Therefore, for high disorder with unscreened interaction, the system decays according to the logarithmic law, but the relaxation time ($\tau = 1/\lambda_{min}$) depends on the degree of disorder and interaction.  	
	The results for the strong screening scenario (d=0.5), when the disorder and interaction are doubled, are also discussed. The lower value of DOS {around the Fermi level} and increase in the width of the distribution in the $W = 4$ case compared to the $W = 2$ case leads to slower relaxation as the disorder increases.  
	
	\section{Discussion} 
	\label{discuss}
	
	In summary, this study discusses the relaxation dynamics of a Coulomb glass model in relation to disorder and screening. The results show that the system relaxes via logarithmic decay at intermediate intervals, regardless of the degree of disorder, when unscreened Coulomb interactions are present. However, the system decays via power law $(\delta n(t) \sim t^{\alpha})$ at late times with a smaller exponent ($\alpha$) as the disorder increases, and for strong disorders, the exponent can be almost zero, with logarithmic decay potentially visible in experiments. The relaxation is faster, and the deviation from logarithmic decay can be observed when the interactions are screened, and the density of states near the Fermi level plays a crucial role in explaining these findings. The system begins to relax gradually when there are few electronic states close to the Fermi level. The depletion of electronic states around the Fermi level is accelerated by increasing disorder and opening a Coulomb gap, which causes slower relaxation. The Coulomb gap is filled and narrowed due to the Coulomb interaction being screened, leading to faster relaxation. In the case of strong disorders with weak screening (where the gap is filled by $10-25\%$), we observe that the relaxation behavior closely resembles that of the unscreened case. These findings are consistent with experimental observations where a metal plate was used to screen the Coulomb interaction, resulting in a filling of the gap by approximately $12-23\%$. Our findings concerning strong screening within the high disorder regime indicate that the relaxation dynamics are primarily governed by the disorder strength. In this scenario, the time required to enter the exponential decay regime decreases compared to the unscreened case, with the main determining factor being the disorder.
	
	Furthermore, the study investigates the separate roles of disorder and interaction strength in determining the relaxation dynamics. In experiments on thin films, increasing the concentration ($n$) of sites in the material increases the interaction between electrons. When $n$ increases, the interaction grows while the average distance between sites ($r_{avg}$) decreases. In a system with strong localization, the disorder strength will be of the order of Fermi energy, which rises with $n$. As a result, an increase in site density causes the ratio of interaction strength ($J = 1/r_{avg}$) to disorder strength ($W$) to decrease by $n^{1/2}$. In our model, a rise in disorder causes a fall in the height and an increase in the width of the density of states, whereas a rise in interaction causes a rise in the width of the Coulomb gap and the DOS. Hence, both disorder and interaction strength contribute equally to sluggish relaxation when they are of similar strength. The relaxation time ($\tau$), after which fluctuations follow the exponential decay law towards equilibrium, is another crucial {parameter in experiments}. In agreement with experiments, our results show that the $\tau$ increases very fast as the strength of the interaction is increased, keeping the disorder strength constant. The increase in $\tau$ is even more when both disorder and interaction strength increase which was also observed experimentally. According to our data, $\tau$ decreases quickly as system disorder decreases, but interaction strength stays the same. This could explain why slow relaxation is not seen in semiconductors with light doping. 
	
	Overall, our results suggest that the intermediate and long-time dynamics are dictated by the DOS near the Fermi energy and far from the Fermi energy, respectively. Thus, and since disorder dictates a pseudogap in the DOS, we find that at large disorders, the decay is logarithmic and slows down with increased disorder. And also, interactions decrease the DOS near the Fermi energy and cause slower dynamics.
		
	\section*{Acknowledgement} 
	Vikas Malik acknowledges the funding from SERB, Department of Science and Technology, Govt of India under the research grant no: CRG/2022/004029. M.S. acknowledges support from the Israel Science Foundation (Grant No. 2300/19). P.B. acknowledges Ben-Gurion University of the Negev for access to their HPC resources. Illuminating discussions with Z. Ovadyahu is gratefully acknowledged. 
	
%

\end{document}